\documentstyle[12pt]{article}
\title{ Evaluation of vacuum energy for tensor fields on spherical spaces}
\author{\large ${E.~N.~Bukina}^{\dagger}$ and N.~N.~Shtykov \\[3mm]
\em Department of Theoretical Physics,
 The Irkutsk State University \\
\em $^{\dagger}$N.~N.~Bogolubov LTF, JINR, 141980, Dubna, Russia}
\date{31 May 1996}
\begin{document}
\maketitle
\begin{abstract}
 The effective one-loop potential on $R^{m+1}\times S^N$ spaces for
massless tensor fields is evaluated. The Casimir energy is given as a value
of $\zeta-$ function by means of which  regularization is made. In even-
dimensional spaces the vacuum energy contains divergent terms coming from poles
of $\zeta(s,q)$ at $s=1$, whereas in odd-dimensional spaces it becomes  
finite.
\end{abstract}
The appearance of divergent vacuum energy is one of peculiarities of quantum
field theory. It is the well-known fact. However, in the presence of
external constraints (external fields, nontrivial topology, curved space) the
absolute value of vacuum energy, in principle,  appears to be a measurble
quantity. For instance, the phenomenon of nontrivial vacuum energy underlies 
the Casimir effect. Vacuum fluctuations in quantum field theory have been
a subject of extensive research. \cite{Plun}
The contribution of vacuum energy in the Robertson-Walker model and Kaluza-Klein
theory, where in place of an external field, the nontrivial background geometry
(curved space) acts as the source to perturb vacuum, was computed in a series
of works 
\cite{Cand},\cite{Birm}, \cite{Chang}.
Analogous evaluations \cite{Sht} are performed for massless scalar fields with 
coupling $\xi R$ and a massive  spinor and vector fields on $R^m \times S^N ,\,
R^m \times CP^2,$ where R is the curvature radius, $R^m$ is the m-dimensional
Euclidean space, $S^N$ is the N-dimensional sphere, and $CP^2$ is the 
projective space.\\
In the present work, we have calculated  the one-loop effective potential for 
massless tensor fields on $R^1 \times S^3, \, R^4 \times S^3.$ \\ [3mm]
{\large \bf 1. The One-Loop Potential for Tensor  on $R^{m+1} \times S^N$}\\[3mm]
For the manifold $R^{m+1} \times S^N$ we introduce the metric as
$
g_{AB} = \eta_{AB} + r^2 h_{AB}
$
where 
$$
\eta_{AB} =
\left[ \begin{array}{c|c}
\eta_{ab}&0\\ \hline
0&0\\    \end{array}   \right],    \qquad
h_{AB} = \left[  \begin{array}{c|c}
0&0\\   \hline
0&h_{\alpha \beta}\\  \end{array}  \right] $$
$$ A = 1, \ldots, m+1+N; \qquad  a = 1, \ldots, m+1; \alpha = 1, \ldots, N $$
the equation for eigenvalues of the Lichnerowitz operator is as follows:
\begin{equation}
(\tilde{\Delta}T)_{AB} = -\nabla_c \nabla^c g_{AB}T_{B}^{D} - 2R_{ADBF} T^{DF}
+R_{AD}T_{B}^{D} + R_{BD}T_{A}^{D} = \lambda T_{AB}
\end{equation}
On  $S^N,$ the curvature tensor  and its convolutions are introduced by
\begin{equation}
R_{\alpha \beta \gamma \delta} = r^2 [h_{\gamma \alpha}h_{\delta \beta} - 
h_{\alpha \delta}h_{\gamma \beta}],\quad
R_{\beta \gamma} = (N-1)h_{\beta \gamma}, \quad
R = N(N-1)/r^2.
\end{equation}
Let us expand the field considered harmonically as follows
\begin{equation}
 T_{ab} =\bar{T_{ab}} (x) Y(y),   \quad
T_{a \alpha} = V_{a}(x)Y_{\alpha}(y),  \quad
 T_{\alpha \beta} = S(x) Y_{\alpha \beta}. 
\end{equation}
In terms of the metric tensor, expressions (2) and  expansion
(3),  equation (1) takes the forms:
$$ (\tilde{\Delta}T)_{ab}=(k^2+\lambda_{0}^{2}) T_{ab},\quad
 (\tilde{\Delta}T)_{a \alpha}=(k^2+\lambda_{1}^{2}) T_{a \alpha} ,\quad
 (\tilde{\Delta}T)_{\alpha \beta} = (k^2+\lambda_{2}^{2}) T_{\alpha \beta},$$
where
$$ -\nabla_{\alpha} \nabla^{\alpha}Y = \lambda_{0}^{2}Y,\quad
 (-\nabla_{\beta} \nabla^{\beta}+N-1)Y_{\alpha}=\lambda_{1}^{2} Y,\quad
 (-\nabla_{\alpha} \nabla^{\alpha}+2N) Y_{\alpha \beta} =\lambda_{2}^{2}
Y_{\alpha \beta}.$$
Throughout we set $r$ of sphere $= 1,$ for simplicity.\\[3mm]

{ \bf 1.1 Evaluation of eigenvalues $\lambda_{2}^{2}$ for  tensor fields} \\[3mm]
The eigenvalues and degeneracies for scalar and tensor fields were
determined in Ref. \cite{Sht}. 
 We consider the symmetric traceless second order tensor in the $(N+1)-$ 
dimensional
Euclidean space $\hat{T_{\hat{\alpha} \hat{\beta}}}$ with $$ 
-\nabla_{\alpha}\nabla^{\alpha} \hat{T_{\hat{\alpha} \hat{\beta}}} =0. $$
The symmetric fields on the sphere $S^N = SO(N+1)/ SO(N)$  can be expressed
through sperical harmonics on $SO(N+1)$ with  the components in SO(N):
$$
 -\nabla^2 T_{\alpha \beta}=[c_{2}^{SO(N+1)}(l,p,0,0,\ldots)-c_{2}^{SO(N)}
(2,0,0,\ldots)] T_{\alpha \beta} .
$$
From the general theory of representations it follows that
$$
c_{2}^{SO(N+1)}(l,p,0,0,\ldots) =l(l+N-2) + p(p+N-4),
$$
which leads to
$$
-\nabla^2 T_{\alpha \beta}=[l(l+N-1)+p(p+N-3)-2N] T_{\alpha \beta}.$$
From the orthogonal expansion of tensor $T_{\alpha \beta}$  we obtain
$$ T_{\alpha \beta} = T_{\alpha \beta}^{\bot}+\nabla_{\alpha} V_{\beta}^{\bot}
+ \nabla_{\beta}V_{\alpha}^{\bot} + (2-\frac{2}{N}) \nabla^2 \phi $$
where $V_{\alpha}=V_{\alpha}^{\bot}+ \nabla_{\alpha}\phi, \qquad 
\nabla^{\alpha} T_{\alpha \beta}^{\bot}=0, \qquad \nabla^{\alpha}
T_{\alpha}^{\bot}=0$. Therefore
\begin{equation}
\lambda_{2_{s}}^{2}=l(l+N-1), \quad
\lambda_{2_{\bot}}^{2}=l(l+N-1)+2(N-1),\quad
\lambda_{2_{v}}^{2}=l(l+N-1) + (N - 2).\quad
\end{equation}
The degenerations of eigenvalues $\lambda_{2_{\bot}}^{2}$   are defined by
dimensional representation $(l,2,0,\ldots)$
\begin{equation}
D_{2}(N,l)=\frac{(N+1)(N-2)(l+N)(l-1)(2l+N-1)(l+N-3)!}{2(N-1)!(l+1)!}.
\end{equation}
The degenerations of $(\lambda_{2_{\bot}}^2)_{v}$ and $(\lambda_{2}^{2})_{s}$
are determined by dimensional representations $(l,1,0,0\ldots)$  and
$(l,0,0,\ldots),$ respectively.\\[3mm]
{\large \bf 2.Evalution of the One-Loop Potential (OLP)}\\[3mm]
The initial object to calculate the effective potential is
$ T_{\mu \nu} (\tilde{\Delta}T)^{\mu \nu}, $
where $\tilde{\Delta}$  was defined in Ref. \cite{Bir}.
The formal expression of OLP is given as
$$
V_{eff}^{(1)} = \frac{1}{2\Omega(R^{m+1})} \mbox{Ln det} \tilde{\Delta} ,
$$
where $\mbox{Ln  det} \tilde{\Delta}$ is given by the following expression 
$$
\begin{array}{c}
 \mbox{Ln det} \tilde{\Delta}= \int d^{m+1}k \{ \sum_{n=0}^{\infty} \frac{(m+1)(m+2)}
{2} D_{0}(N,n)\mbox{Ln}(\lambda_{0}^{2}+k^2) 
+ \\[12pt]
+ \sum_{n=1}^{\infty} (m+1) D_1(N,n) Ln(\lambda_{1}^{2} + k^2) + \\[12pt]
+\sum_{n=1}^{\infty} (m+1) D_0(N,n) Ln(\lambda_{0}^{2} + k^2)
+\sum_{n=2}^{\infty} D_1(N,n) Ln(\lambda_{1}^{2} + k^2) + \\[12pt]
+\sum_{n=2}^{\infty} D_2(N,n) Ln(\lambda_{2}^{2} + k^2) + 
\sum_{n=2}^{\infty} D_0(N,n) Ln(\lambda_{0}^{2} + k^2) \} 
\end{array}
$$
 With $D_{i}(N,n)$ and 
$\lambda_{i}^{2}$ the regularized effective OLP has the form 
$$
\begin{array}{c}
V_{tn}^{(1)} = Lim_{s \to 1} \frac{\Gamma(-s-m/2) \mu^{2\epsilon}}
{2(4\pi)^{m/2} \Gamma(-s/2)}[ \sum_{n=0}^{\infty} \frac{(m+1)(m+2)}{2}
\frac{(2n+N-1) \Gamma(n+N-1)}{\Gamma(N) n!}(n(n+N-1))^{\frac{m+s}{2}}+\\[12pt]
+ \sum_{n=1}^{\infty}(m+1)\frac{n(n+N-1)(2n+N-1)(n+N-3)!}{(N-2)!(n+1)!}
(n(n+N-1)+(N-2))^{\frac{m+s}{2}} + \\[12pt]
+\sum_{n=1}^{\infty}(m+1) \frac{(2n+N-1) \Gamma(n+N-1)}{\Gamma(N) n!} 
(n(n+N-1))^{\frac{m+s}{2}} +\\[12pt]
+ \sum_{n=2}^{\infty} \frac{n(n+N-1)(2n+N-1)(n+N-3)!}{(N-2)!(n+1)!}
(n(n+N-1)+(N-2))^{\frac{m+s}{2}}+ \\[12pt]
+ \sum_{n=2}^{\infty} \frac{(2n+N-1)(n+N-3)!(n+N)(N+1)(N-2)(n-1)}{2(N-1)!(n+1)!}
(n(n+N-1)+2(N-1))^{\frac{m+s}{2}}+ \\[12pt]
+ \sum_{n=2}^{\infty} \frac{(2n+N-1) \Gamma(n+N-1)}{\Gamma(N) n!}
(n(n+N-1))^{\frac{m+s}{2}} ]
\end{array}
$$
For $R^1 \times S^3$ representing the Robertson-Walker space-time, 
$ V_{tn}^{(1)}(m=0,N=3) \approx -3.6563\frac{1}{\epsilon} - 4.2372,$
where the relation
$$
\sum_{n=0}^{\infty} (n+q)^{l} ((n+q)^2 +M)^{\frac{m+s}{2}} = 
\sum_{n=0}^{\infty}\frac{\Gamma(k-\frac{m+s}{2})(-1)^{k}}{\Gamma(\frac{-m-s}{2})
k!} \zeta(2k-l-m-s,q)M^{k}
$$
is used. 
Also, we obtained  for $R^4 \times S^3$: $V_{tn}^{(1)}(m=3,N=3)
\approx 0.0262. $
In this case the $\Gamma-$ function poles for negative values of the argument of
$\Gamma (k-2)$ at $k=0,1,2$ are compensated by zero of $\zeta-$function at 
negative integer values of its argument according to the reflection property:
$$
\Gamma(-z)\zeta(-2z)=\frac{\zeta(1+2z) \Gamma(\frac{1+2z}{2})}{\pi^{2z+1/2}}
$$
with $\Gamma(-l) = \Gamma(-k) (-1)^{k+l} k!/l!$.
It should be emphasized that the OLP on odd-dimensional manifolds has  finite
values. This is evident from the spectral properties of the Lichnerowicz operator
that contrubutes only to the kinetic part of the Lagrangian, and the corresponding
invariant counterterms $R^2,\, R_{\mu \nu}^{2}, \, R_{\mu \nu \sigma \delta}^{2}$
contain an even degree of $r$,only. Therefore they can't be constructed in an
odd-dimensional space without some additional  dimensional parameters.
And vice versa, in an even-dimensional space the divergent part may be compensated
by a specific combination of invariants constructed from the curvature tensor
and  contractions of the latter. In a $4-$ space this combination looks as
$
\alpha_1 R^2 + \alpha_2 R_{\alpha \beta}^{2} + \alpha_3 R_{\alpha \beta \gamma
\delta}^{2}$
with  arbitrary coefficients $\alpha_{i}$.
For the case of scalar and spinor fields this combination is defined as a
conformal anomaly and turns out to be \cite{Bir}
$$
\begin{array}{c}
A_{(sc)}= \frac{1}{180}R_{\alpha \beta \gamma \sigma}
R^{\alpha \beta \gamma \sigma} - \frac{1}{180}R_{\alpha \beta}R^{\alpha \beta}
+ \frac{1}{72} R^2,\\ [12pt]
A_{(sp)}= \frac{1}{720} \left[ \frac{5}{2} R^2 -\frac{7}{2}R^{\mu \nu \rho \sigma}
R_{\mu \nu \rho \sigma}-4R^{\mu \nu}R_{\mu \nu} \right].
\end{array}
$$
The coefficients $\alpha_{i}$ were computed in a 4-dimensional  space for the
electromagnetic and gravitational fields \cite{Toms}.
The results demonstrate that the values of OLP for tensor
fields give the main contribution to the vacuum energy in comparison with 
scalar, spinor and vector fields ( compare with results of Ref. \cite{Sht}).
This means that the contributions to the 
vacuum energy on a compact space are ensured by the 
quantum corrections due to the gravitational field.

\end{document}